\documentclass[a4paper,fleqn]{cas-sc}

\usepackage[authoryear,longnamesfirst]{natbib}
\usepackage{wrapfig} 
\usepackage{flushend} 
\usepackage{float}
\usepackage{hyperref}
\usepackage{url}
\usepackage{times}
\usepackage{latexsym}
\usepackage{times}
\usepackage{soul}
\usepackage{url}
\usepackage{amssymb} 
\usepackage{multirow}
\usepackage{amsthm}
\usepackage{amsmath}
\usepackage{booktabs}
\usepackage{algorithm}
\usepackage{algorithmic}
\usepackage[switch]{lineno}
\usepackage[T1]{fontenc}
\usepackage[utf8]{inputenc}
\usepackage{microtype}
\usepackage{inconsolata}
\usepackage{graphicx}
\usepackage{tabularx,booktabs}

\usepackage{subfig}
\usepackage{listings, xcolor}
\definecolor{verylightgray}{rgb}{.97,.97,.97}

\lstdefinelanguage{Solidity}{
	keywords=[1]{anonymous, assembly, assert, balance, break, call, callcode, case, catch, class, constant, continue, constructor, contract, debugger, default, delegatecall, delete, do, else, emit, event, experimental, export, external, false, finally, for, function, gas, if, implements, import, in, indexed, instanceof, interface, internal, is, length, library, log0, log1, log2, log3, log4, memory, modifier, new, payable, pragma, private, protected, public, pure, push, require, return, returns, revert, selfdestruct, send, solidity, storage, struct, suicide, super, switch, then, this, throw, transfer, true, try, typeof, using, value, view, while, with, addmod, ecrecover, keccak256, mulmod, ripemd160, sha256, sha3}, 
	keywordstyle=[1]\color{blue}\bfseries,
	keywords=[2]{address, bool, byte, bytes, bytes1, bytes2, bytes3, bytes4, bytes5, bytes6, bytes7, bytes8, bytes9, bytes10, bytes11, bytes12, bytes13, bytes14, bytes15, bytes16, bytes17, bytes18, bytes19, bytes20, bytes21, bytes22, bytes23, bytes24, bytes25, bytes26, bytes27, bytes28, bytes29, bytes30, bytes31, bytes32, enum, int, int8, int16, int24, int32, int40, int48, int56, int64, int72, int80, int88, int96, int104, int112, int120, int128, int136, int144, int152, int160, int168, int176, int184, int192, int200, int208, int216, int224, int232, int240, int248, int256, mapping, string, uint, uint8, uint16, uint24, uint32, uint40, uint48, uint56, uint64, uint72, uint80, uint88, uint96, uint104, uint112, uint120, uint128, uint136, uint144, uint152, uint160, uint168, uint176, uint184, uint192, uint200, uint208, uint216, uint224, uint232, uint240, uint248, uint256, var, void, ether, finney, szabo, wei, days, hours, minutes, seconds, weeks, years},	
	keywordstyle=[2]\color{teal}\bfseries,
	keywords=[3]{block, blockhash, coinbase, difficulty, gaslimit, number, timestamp, msg, data, gas, sender, sig, value, now, tx, gasprice, origin},	
	keywordstyle=[3]\color{violet}\bfseries,
	identifierstyle=\color{black},
	sensitive=true,
	comment=[l]{//},
	morecomment=[s]{/*}{*/},
	commentstyle=\color{orange}\ttfamily,
	stringstyle=\color{red}\ttfamily,
	morestring=[b]',
	morestring=[b]"
}

\lstdefinelanguage{json}{
    basicstyle=\normalfont\ttfamily,
    numbers=left,
    numberstyle=\scriptsize,
    basicstyle=\footnotesize\ttfamily,
    breaklines=true,
    showstringspaces=false,
    string=[db]{"},
    stringstyle=\color{green!50!black},
    morestring=[s][\color{black}]{\ \ "}{":},
    keywordstyle=\color{blue},
    keywords={true,false,null},
    literate=
     *{0}{{{\color{red}0}}}{1}
      {1}{{{\color{red}1}}}{1}
      {2}{{{\color{red}2}}}{1}
      {3}{{{\color{red}3}}}{1}
      {4}{{{\color{red}4}}}{1}
      {5}{{{\color{red}5}}}{1}
      {6}{{{\color{red}6}}}{1}
      {7}{{{\color{red}7}}}{1}
      {8}{{{\color{red}8}}}{1}
      {9}{{{\color{red}9}}}{1}
      {.}{{{\color{red}.}}}{1}
      {:}{{{\color{gray}{:}}}}{1}
      {,}{{{\color{gray}{,}}}}{1}
      {\{}{{{\color{gray}{\{}}}}{1}
      {\}}{{{\color{gray}{\}}}}}{1}
      {[}{{{\color{gray}{[}}}}{1}
      {]}{{{\color{gray}{]}}}}{1},
}

\def\tsc#1{\csdef{#1}{\textsc{\lowercase{#1}}\xspace}}
\tsc{WGM}
\tsc{QE}
\tsc{EP}
\tsc{PMS}
\tsc{BEC}
\tsc{DE}

\begin{document}
\let\WriteBookmarks\relax
\def\floatpagepagefraction{1}
\def\textpagefraction{.001}

\shorttitle{Smart Contract Library Misuse Detection with Iterative Feedback and Static Verification}

\shortauthors{Yishun Wang et~al.}

\title [mode = title]{LibScan: Smart Contract Library Misuse Detection with Iterative Feedback and Static Verification}     


\author[1]{Yishun Wang}[orcid=0009-0003-1525-3393]

\fnmark[1]

\ead{yishunwang@hainanu.edu.cn}


\credit{Conceptualization, Data curation, Methodology, Software, Writing – original draft}

\affiliation[1]{organization={Hainan University},
    city={Haikou},
    postcode={570228}, 
    country={China}}

\author[1]{Wenkai Li}[orcid=0000-0002-4238-8846]
\fnmark[1]
\ead{cswkli@hainanu.edu.cn}
\credit{Conceptualization, Data curation, Formal analysis, Investigation, Validation, Writing – review \& editing}



\author[1]{Xiaoqi Li}[orcid=0000-0002-6012-9178]
\ead{csxqli@ieee.org}

\cormark[1]
\credit{Funding acquisition, Project administration, Resources, Supervision}

\author[1]{Zongwei Li}[orcid=0009-0008-7233-9464]
\ead{lizw1017@hainanu.edu.cn}
\credit{Validation, Writing – review \& editing}

\author[1]{Lei Xie}[orcid=0009-0007-2883-206X]
\ead{xielei@hainanu.edu.cn}
\credit{Validation, Visualization}

\author[1,2]{Yuqing Zhang}[orcid=0000-0001-8306-7195]
\ead{zhangyq@nipc.org.cn}

\affiliation[2]{organization={University of Chinese Academy of Sciences},
    city={Beijing},
    postcode={100049}, 
    country={China}}
\credit{Resources, Supervision}

\cortext[cor1]{Corresponding author}

\fntext[fn1]{Yishun Wang and Wenkai Li contributed equally to this work.}

\begin{abstract}
Smart contracts are self-executing programs that manage financial transactions on blockchain networks. Developers commonly rely on third-party code libraries to improve both efficiency and security. However, improper use of these libraries can introduce hidden vulnerabilities that are difficult to detect, leading to significant financial losses. Existing automated tools struggle to identify such misuse because it often requires understanding the developer's intent rather than simply scanning for known code patterns.
This paper presents LibScan, an automated detection framework that combines large language model (LLM)-based semantic reasoning with rule-based code analysis, identifying eight distinct categories of library misuse in smart contracts. To improve detection reliability, the framework incorporates an iterative self-correction mechanism that refines its analysis across multiple rounds, alongside a structured knowledge base derived from large-scale empirical studies of real-world misuse cases.
Experiments conducted on 662 real-world smart contracts demonstrate that LibScan achieves an overall detection accuracy of 85.15\%, outperforming existing tools by a margin of over 16 percentage points. Ablation experiments further confirm that combining both analysis approaches yields substantially better results than either method used independently.


\end{abstract}




\maketitle

\section{Introduction}
With the booming development of the Ethereum and Web3 ecosystems~\citep{vaigandla2023review}, smart contracts have become the core infrastructure of decentralized finance (DeFi)~\citep{kapengut2023event}, hosting and managing hundreds of billions of dollars in crypto assets. However, the complete openness of their code and the immutability of the blockchain make them a prime target for various cyberattacks. Extremely minor flaws in the code logic can often lead to irreversible and catastrophic financial losses. For example, the Parity multi-signature wallet vulnerability~\citep{ParityHack2017OpenZeppelin}, caused by improper initialization and arbitrary calls to the underlying library contract, ultimately resulted in the permanent locking of over \$150 million worth of Ethereum. Recent security statistics further demonstrate that attacks targeting smart contract design flaws still cause hundreds of millions of dollars in losses annually~\citep{chu2023survey,zheng2024dappscan}. This not only highlights the importance of smart contract vulnerability discovery and security auditing but also compels the development of more precise security detection mechanisms.

To reduce the risk of redundant development and improve code standardization, developers are increasingly relying on Solidity libraries (e.g., SafeERC20 and SafeMath) when writing smart contracts~\citep{sharma2023review}. A library is essentially a set of reusable functions, typically used to encapsulate common business logic or address specific security vulnerabilities~\citep{huang2024revealing}. Theoretically, introducing rigorously tested and audited libraries can significantly improve the security baseline of smart contracts~\citep{zhang2022cbgru}. However, due to information asymmetry between library creators, contract developers, and community maintainers, the hidden and dangerous vulnerability of library misuse is gradually spreading in the blockchain ecosystem. Such improper behaviors~\citep{huang2024revealing} mainly contain Invalid Wrapper Check in Library, Unhandled Exceptions in Library, Inappropriate Library Extension, Inappropriate Using For, Incomplete Function Replacement, Overestimated Library Capability, Underestimated Library Capability, and Unnecessary Library Using.

While smart contract vulnerability detection has been extensively studied, automated detection of the specific interdisciplinary area of library misuse still faces significant technical bottlenecks. Traditional static analysis tools (e.g., Slither~\citep{Slither2024} and Smartbugs~\citep{di2023smartbugs}) are widely used in industrial security scanning due to their excellent scalability. However, these tools heavily rely on hard-coded rules, abstract syntax trees (ASTs), or control flow graphs (CFGs), often proving inadequate when dealing with misuse patterns that require a deep understanding of the codebase's design intent, developer subjective assumptions, and complex contextual semantics. On the other hand, the emergence of large language models (LLMs)~\citep{sun2024gptscan} has demonstrated revolutionary potential in code semantic understanding and cross-language logical reasoning. Previously, researchers attempted to use generative models such as GPT~\citep{david2023you} to directly diagnose contract vulnerabilities. However, due to the lack of strict code-level constraints, purely data-driven methods are prone to severe illusion phenomena when dealing with long and complex smart contracts, resulting in extremely high false positive rates~\citep{david2023you}.

Based on the above challenges, existing smart contract security systems lack a unified detection framework that combines high-level semantic understanding with precise code-level verification. To address this gap, this paper proposes LibScan, a hybrid detection framework designed for smart contract library misuse auditing.
LibScan integrates the semantic reasoning capabilities of large language models with the deterministic precision of static analysis techniques, including structural matching via TF-IDF matrices and customized Slither rule sets. This combination allows the framework to leverage the strengths of both approaches while compensating for their individual limitations.
To further improve reliability, LibScan incorporates a closed-loop iterative feedback mechanism and a pattern-aware knowledge base constructed from large-scale empirical studies. Moreover, these components enable the framework to align contextual information more accurately, reduce hallucinations produced by language models, and lower false positive rates in detection results. To enhance the reliability of GPT's detection capabilities, we have implemented a PTS prompting strategy following extensive testing on diverse contracts. Drawing inspiration from the Auto-CoT \citep{kashun2023automatic} prompt and the LTS (i.e., Let's think step by step) strategy of zero-shot-CoT \citep{li2023prompt}, this technique encourages the LLM to articulate its reasoning process. The resulting rationale and questions are subsequently integrated into a guiding prompt, steering the LLM toward generating responses. The main contributions are as follows:

\begin{itemize}
    \item We propose LibScan, to our best knowledge, which is the first tool to use LLM combined with static analysis to detect misuse patterns in smart contract libraries.
    \item We integrate descriptions of various library misuse patterns in smart contracts along with code-level scenarios. This integration provides an excellent matching standard for using LLM to detect library misuse patterns.
    \item LibScan has shown satisfactory results in detecting library misuse patterns across a comprehensive sample of 183 real-world smart contracts, and it has also achieved an acceptable level of detection for complex contracts.
    \item We have uploaded the related codes and experimental data of LibScan at \url{https://figshare.com/s/147dbeb413c027806674}.
\end{itemize}

\section{Background}

\subsection{Ethereum and Smart Contracts}
Ethereum \citep{tikhomirov2017ethereum,kado2025empirical}, launched in 2015, revolutionized the blockchain landscape by introducing a programmable blockchain that supports the creation of smart contracts. Smart contracts \citep{lin2022survey,bartoletti2025smart,negi2026smart} are self-executing contracts with the terms of the agreement directly written into lines of code, running on the blockchain. These contracts hold the potential to digitize and streamline a vast array of processes across various industries.

The Ethereum platform allows developers to write smart contracts using its leading language, Solidity, and deploy them on the Ethereum Virtual Machine (EVM) \citep{hirai2017defining,campos2025analise,zhu2024sybil,wu2025security}. The EVM is a decentralized virtual machine that executes code using an international network of public nodes. This approach brings about a higher degree of transparency and trust, as transactions and contract states are recorded immutably on the blockchain.
\subsection{Solidity Library}
Solidity libraries \citep{palechor2022solidity,yang2025multi,li2025beyond} are a collection of functions that are reusable and can be called by other contracts. Developers often encapsulate common operations or functions that solve specific problems. When other developers wish to implement similar functionalities, they can import these libraries and utilize their methods to address particular issues~\citep{bu2025smartbugbert,zhang2025penetration}. In Solidity, libraries are defined by the keyword "library" and are linked to the contract during compilation. Libraries typically facilitate code reuse, a concept that is present in many mainstream programming languages today, such as Java, Python, C++, etc.
\subsection{Library Misuse}
Library misuse is a very common issue in developing smart contracts, involving three key roles: the developers who create the libraries, the contract developers who use the libraries, and the community that maintains the libraries \citep{huang2024revealing}. Effective collaboration among these three roles is conducive to developing smart contracts. However, improper behavior by the developers who create the libraries or the contract developers who use them may lead to unintended negative consequences. Such improper behavior is known as library misuse~\citep{wu2025exploring,ding2025comprehensive}. It includes but is not limited to incorrect function calls, a lack of full understanding of the capabilities of library functions, or failure to update to a secure version of the library promptly, thereby introducing security vulnerabilities, among other things.

We have consolidated the attributes of eight types of library misuse patterns along with code-level scenarios, which will be thoroughly explained in the section Preliminaries.

\section{Preliminaries}
In this section, we detail eight potential library misuse patterns in the development and management of smart contracts. Owing to space limitations, we present code examples for a select few patterns identified as prevalent in the empirical research. These eight patterns were first introduced in the empirical research by \cite{huang2024revealing}, and we have integrated them to become the main data source for this paper.
\subsection{P1: Invalid Wrapper Check in Library}
\textbf{Pattern introduction: }The logic implemented in library functions for checks is flawed, failing to handle all expected scenarios, leading to security issues correctly. For example, SafeERC20 \citep{parisi2024defi} is a library designed to safeguard ERC-20 token operations. Improper use of this library by developers, such as failing to check for return values, can lead to potential risks. 

Listing \ref{lst:Listing1} illustrates a flaw in the approve function where a contract cannot update the allowance by invoking the function with a non-zero value. If the spender has already utilized the allowance from an initial approval, a subsequent approval does not change the original allowance but adds a new one. To address this, the safeApprove function replaces the approve function, enabling a contract to set the allowance to zero or modify it from zero, thus preventing an unauthorized direct reset of the allowance, which is in line with the library's intended functionality.

\begin{lstlisting}[language=Solidity,caption={Invalid Wrapper Check in Library},
  label={lst:Listing1}]
library SafeERC20 {
    function safeApprove(IERC20 token, address spender, uint256 value) internal {
        require((value == 0) && (token.allowance(msg.sender, spender) == 0));
        require(token.approve(spender, value));
}}
\end{lstlisting}
\textbf{Optimization and mitigation: }Developers should (i) rigorously validate wrapper logic against formal specifications, (ii) prefer well-audited and widely adopted library versions, and (iii) avoid modifying wrapper semantics without comprehensive regression testing. From a tooling perspective, static analysis should verify that wrapper conditions are both necessary and sufficient with respect to the intended safety guarantees.

\subsection{P2: Unhandled Exceptions in Library}
\textbf{Pattern introduction: }The library was not designed with consideration for all possible exceptions, leading to contract execution failure under certain circumstances \citep{khan2022code}. For instance, a library function may interact with other contracts during execution, but does not handle the exceptional responses from those contracts. As shown in Listing \ref{lst2}.

\begin{lstlisting}[language=Solidity,caption={Inappropriate Library Extension},label={lst2}]
function safeTransfer(address to, uint256 amount) internal {
(bool success, ) = to.call(abi.encodeWithSignature("receive(uint256)", amount));
// Missing explicit exception handling
require(success, "Transfer failed");
}
\end{lstlisting}
If $success$ is not checked or decoded correctly, the contract may fail to detect execution errors.
\textbf{Optimization and mitigation: }Libraries should explicitly validate return values and revert on failure. Developers are advised to use high level interfaces when possible and ensure that all external interactions are guarded by explicit success checks. Automated tools should flag unchecked low-level calls within library code.

\subsection{P3: Inappropriate Library Extension}
\textbf{Pattern introduction: }Multiple functionalities are mixed within library functions, making them overly complex, difficult to maintain and understand, and increasing the risk of errors. For example, a library function is responsible for both calculating and verifying several different business logics, increasing the likelihood of mistakes.

As illustrated in Listing \ref{lst:Listing2}, the ECDSA \citep{yang2022efficient} is utilized to recover the address from a signature. The recover function within the library can recover signatures that include r, s, and v, where v is a boolean value limited to either the 27/28 version format or the 0/1 version format, with the 27/28 version being the most commonly used. Initially, the recover function only accepted the 27/28 version as input at line six. To extend support for the 0/1 version, a simple conditional logic was added at line five within the recover function, rather than creating a new function dedicated to supporting the 0/1 version format. This introduces a potential security risk.

\begin{lstlisting}[language=Solidity,caption={Inappropriate Library Extension},label={lst:Listing2}]
library ECDSA{
    function recover(bytes32 hash , bytes memory signature)
internal pure returns (address){
    bytes32 r; bytes32 s; uint8 v;
    // Code omitted: Calculating r, s, v from signature
    if (v < 27) { v += 27;} // ERROR
}}
\end{lstlisting}
\textbf{Optimization and mitigation: }Library designers should introduce separate functions for distinct behaviors and deprecate legacy interfaces gradually. Clear documentation and versioning are essential to prevent misuse by downstream contracts.

\subsection{P4: Inappropriate Using For}
\textbf{Pattern introduction: }The 'using for' statement is incorrectly used in contracts, binding library functions to incompatible or unnecessary data types. For instance, the SafeMath \citep{zhou2023security} library can be applied to the int256 type when it is designed exclusively for uint256. As shown in Listing \ref{lst4}.

\begin{lstlisting}[language=Solidity,caption={Inappropriate Library Extension},label={lst4}]
using SafeMath for int256; // SafeMath is designed for uint256
int256 a = -1;
int256 b = a.add(1); // Semantically incorrect usage
\end{lstlisting}
\textbf{Optimization and mitigation: }Developers should ensure that using for bindings strictly match the intended data types of the library. Static analyzers can detect mismatches between library design assumptions and actual bindings.

\subsection{P5: Incomplete Function Replacement}
\textbf{Pattern introduction: }In contracts, the original unsafe functions are not fully replaced with safer versions provided by the library. For example, developers, when using the SafeERC20 library, fail to replace all approved calls with safeApprove, missing some replacements.

\textbf{Pattern introduction: }Once a safer alternative is adopted, all legacy calls should be systematically replaced. Automated refactoring tools and static checks can help ensure consistency across the codebase.

\subsection{P6: Overestimated Library Capability}
\textbf{Pattern introduction: }Developers wrongly assume that library functions have certain capabilities that are not implemented by the library. For example, using a library function to check if an address is a contract, but the function does not verify the contract's deployment code \citep{khan2024defi}.

As shown in listing \ref{lst:Listing3}, the capacity of the contract function in the BondingCurve contract is overstated because it only checks whether the length of the executing code is empty, without considering the length of the constructor. Therefore, the $isContract$ function on line 3 cannot be used to reject contracts that only have a constructor.

\begin{lstlisting}[language=Solidity,caption={Overestimated Library Capability},label={lst:Listing3}]
contract BondingCurve{
    function allocate () external {
        require ((! Address.isContract(msg.sender)) || msg.sender == core().genesisGroup (), 
        "BondingCurve:Caller is a contract"); // ERROR
        // ...
}   }
\end{lstlisting}
\textbf{Pattern introduction: }Developers should carefully review library documentation and underlying implementations. Critical security checks should not rely on a single heuristic but instead combine multiple defensive mechanisms.
\subsection{P7: Underestimated Library Capability}
\textbf{Pattern introduction: }Developers fail to fully utilize the functionalities provided by library functions, leading to the redundant implementation of functions already offered by the library within the contract. For instance, developers manually implement a calculation function in the contract when the library already provides an optimized implementation. As shown in Listing \ref{lst7}.

\begin{lstlisting}[language=Solidity,caption={Overestimated Library Capability},label={lst7}]
function safeAdd(uint256 a, uint256 b) internal pure returns (uint256) {
uint256 c = a + b;
require(c >= a, "Overflow");
return c;
}
\end{lstlisting}
When $SafeMath$ already provides equivalent functionality, this implementation is unnecessary.
\textbf{Pattern introduction: }Developers should perform a thorough review of available library APIs before implementing custom logic. Tooling support can detect redundant implementations that duplicate known library functions.
\subsection{P8: Unnecessary Library Using}
\textbf{Pattern introduction: }Unnecessary libraries are included in contracts, which are either not used or already natively supported by newer versions of the compiler, leading to resource waste. For example, even though Solidity 0.8 has built-in overflow checks, developers still introduce the SafeMath library for overflow protection, causing unnecessary gas consumption.

\begin{figure}
    \centering
    \includegraphics[width=0.80\linewidth]{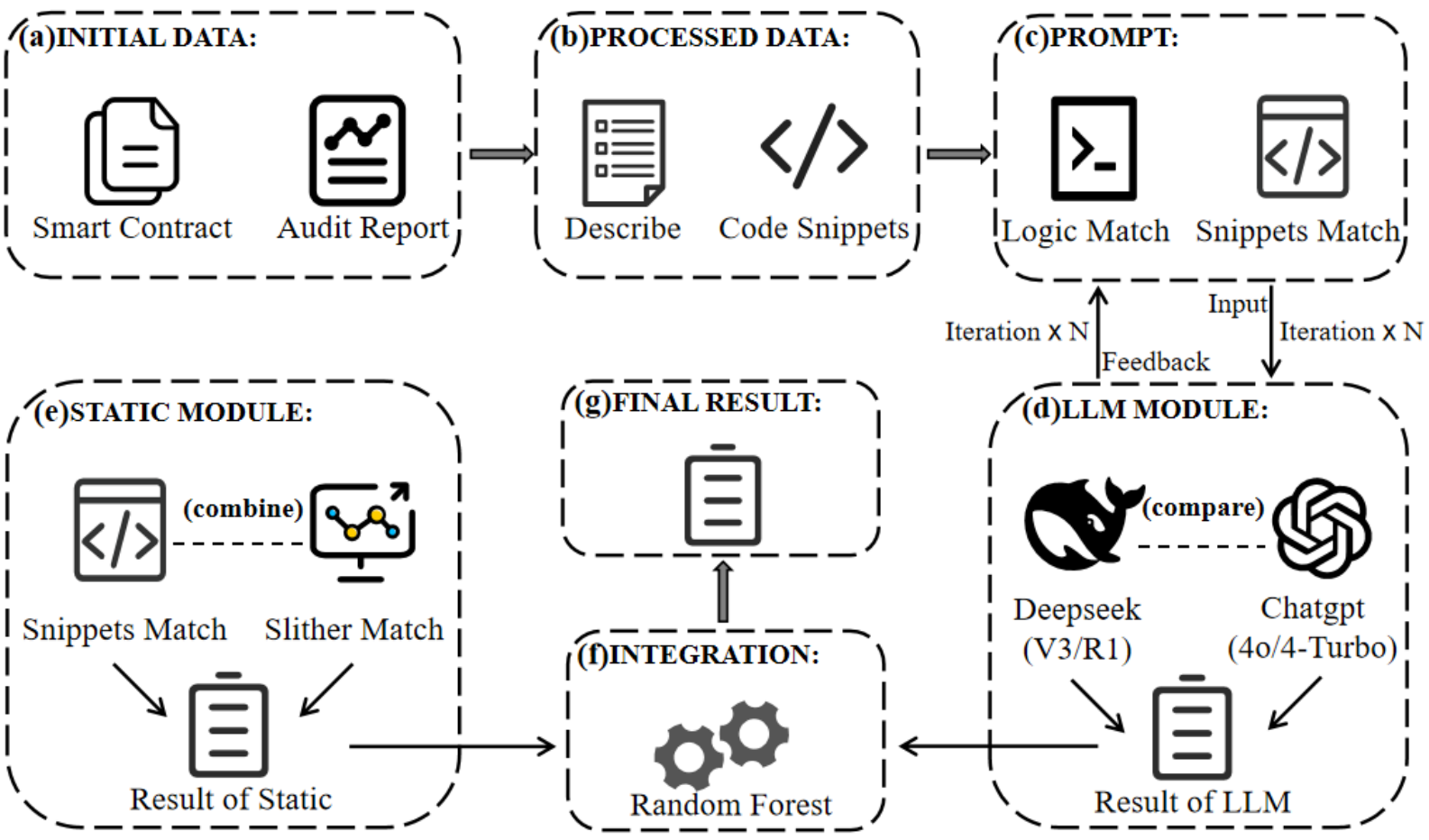}
    \caption{Overall architecture of LibScan, a hybrid framework for smart contract library misuse detection. LibScan combines dual heterogeneous LLM-based semantic reasoning with static analysis and a loop feedback optimization mechanism to reconcile inconsistent results, suppress hallucinations, and incrementally update a pattern-aware knowledge base derived from empirical misuse studies.}
    \label{fig:1}
\end{figure}

\textbf{Pattern introduction: }Developers should align library usage with the target compiler version and periodically refactor dependencies. Static analysis tools can identify unused imports and recommend their removal to reduce deployment and execution costs.

\begin{lstlisting}[language=json,caption={Example of JSON in P2 pattern},label={lst:Listing4}]
{
    "pattern": "P2",
    "name": "Unhandled Exceptions in Library",
    "description": "This misuse pattern occurs when the library lacks exception handling. During library development, the library developer may not consider the robustness and security requirements in Solidity, and only implement the core business feature.",
    "snippets": [
        "return success && abi.decode(result, (bool));",
        "return (success && result.length == 32 && abi.decode(result, (bytes4)) == IERC1271.isValidSignature.selector);"
      ]},
\end{lstlisting}

\section{Methodology}
In this section, we detail the primary process of LibScan, as depicted in Figure \ref{fig:1}.

Initially, we manually integrated empirical research data to construct scenario descriptions and code properties for eight library misuse patterns. listing \ref{lst:Listing4} uses the P2 pattern as an example and presents the pattern name, structured scenario description, and code properties in JSON format. Subsequently, upon receipt of a Solidity smart contract code file, LibScan initiates the parsing process via the LLM module. The LLM module employs GPT to parse the contract code, integrating feature matching of misuse patterns with logical inferences drawn from scenario descriptions, thereby identifying the misuse patterns and associated code segments within the input contract. This output is not the LLM module's outcome; it serves as input for subsequent iterative GPT parsing prompts. Following this, a static analysis is applied to the contract, and ultimately, the findings from both modules are synthesized to yield conclusive results.

\subsection{LLM-Based Contract Library Misuse Pattern Detection}
\textbf{Prompt for LLM.}We define library misuse patterns as $M_{pattern} \in \{ P1, P2, P3, P4, P5, P6, P7, P8, NONE \}$, $P1$ to $P8$ indicate the corresponding library misuse patterns present in the Solidity contract, whereas “$NONE$” indicates the absence of any such misuse patterns in the contract. We have consolidated the names of the 8 library misuse patterns, attribute descriptions for each pattern, and code characteristics. The attribute descriptions serve as the basis for LLM's logical deduction of contract code, while the code characteristics are used for LLM's code-level matching of the input contract. 

Since LLM's outputs are not limited to binary outcomes, "$NONE$" is categorized as the negative class, and P1 to P8 are categorized as the positive classes for metric calculations. With these classifications, we determine metrics including accuracy, recall, F1 score, and precision. A confusion matrix can be constructed to assess the performance of the output results:

If the $predict\_pattern$ from the output matches the $true\_pattern$ of the input contract and $true\_pattern$ is not the negative class $NONE$ the result is a True Positive (TP). Conversely, if the $true\_pattern$ is $NONE$ the output is a True Negative (TN). When the $predict\_pattern$ does not align with the $true\_pattern$ and $true\_pattern$ is not $NONE$ the result is a False Negative (FN). If the $true\_pattern$ is $NONE$ the result is a False Positive (FP).

\begin{figure}
\centering
\includegraphics[width=0.65\columnwidth]{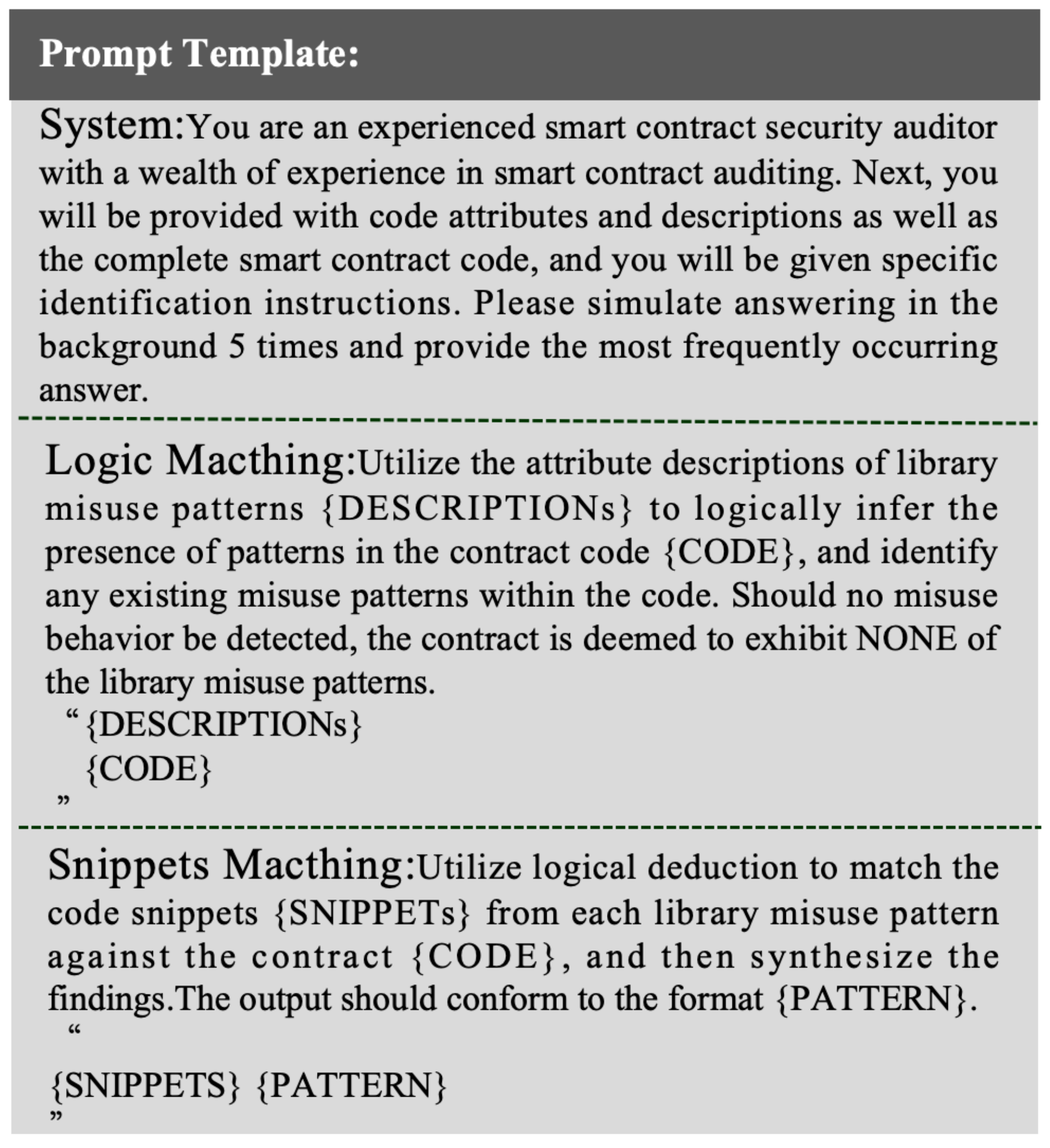} 
\caption{Prompt for Logic and Snippets Matching}
\label{fig2}
\end{figure}

To enhance LLM-generated responses' reliability and mitigate the impact of complex contracts on its code reasoning and feature matching capabilities, we introduce an innovative PTS (i.e., Pre-generated, Then Selected) prompting strategy for LLM, inspired by the Auto-CoT \citep{paranjape2023art} framework. Initially, we categorize the query into distinct clusters, each corresponding to one of the eight library misuse behavior patterns within the LibScan context. Subsequently, adopting the reflective approach LTS (i.e., Let's think step by step) from Zero-shot-CoT, we employ a prompt to induce a series of analytical thought processes from the LLM. These rationales are then integrated with the question and re-presented with an emphasis on the answer, as illustrated in Figure \ref{fig2}. By instructing the LLM system to iterate the question-response process five times and report the most frequent response, we can significantly reduce the inherent indeterminacy of the answers.

\textbf{Feedback for Achieving Iterative Optimization of LLM.} 
Our directives for LLM encompass matching code features and engaging in logical deduction from attribute descriptions, necessitating robust capacities for both logical reasoning and computational linguistics. Consequently, GPT-3.5 is deemed unsuitable. We therefore employ a combination of GPT-4 \citep{sonoda2024diagnostic}, GPT-4 Turbo \citep{wada2024prompt}, DeepSeek-R1 \citep{neha2025survey}, and DeepSeek-V3 \citep{liu2024deepseek} models - all demonstrating enhanced capabilities in symbolic reasoning and semantic parsing. The DeepSeek series particularly excels in formal specification processing through their hybrid architecture combining 192B parameters with retrieval-enhanced training. GPT-4 Turbo handles up to 128k token contexts through its compressed attention mechanism, enabling precise processing of lengthy technical documents with temporal consistency. While GPT-4o matches Turbo's performance on English code/text processing, DeepSeek-V3 demonstrates comparable reasoning accuracy while achieving 3.2× faster inference speeds through dynamic sparse computation. For multimodal interaction and non-English text processing, GPT-4o remains optimal, whereas DeepSeek-R1 shows particular efficacy in cross-lingual code comment alignment through its contrastive pretraining framework.

The LLM model's outputs are inherently random \citep{kikuchi2024towards}, meaning it can provide different results for the same contract code. To reduce this variability, we set the temperature parameter to zero when using GPT. Lowering this parameter in a language model decreases randomness and increases predictability. We randomly selected 200 contracts from 1,018 manually inspected and labeled contracts as test data, following these principles:
\begin{itemize}
\item \textbf{Diversity}: The sample should include a variety of library misuse patterns, including "$NONE$".
\item \textbf{Balance}: The ratio of positive to negative classes in the sample should be consistent with that of the overall dataset.
\item \textbf{Randomness}: Use random sampling methods to reduce sampling bias.
\end{itemize}
We applied an identical prompt to the test data for all models, ensuring consistency in all other parameters. All models produce data in a uniform format, detailing the type of library misuse pattern, its name, a descriptive attribute, and the corresponding code snippet illustrating the misuse. Subsequently, we calculated four evaluation metrics: accuracy, precision, recall, and the F1 score using the binary classification confusion matrix described earlier and in accordance with the rules outlined below.

$Accuracy = \frac{1}{N}\sum_{i=1}^N\mathbb{I}(\hat{y}_i = y_i)$,
 where: $N$ denotes the total number of samples, $\hat{y}_i$ is the predicted class for the i-th sample, while $y_i$ represents the true class, and $\mathbb{I}$ is an indicator function which equals 1 when the prediction is correct and 0 otherwise.
 
$Precision = \frac{1}{K}\sum_{i=1}^K\frac{TP_k}{TP_k+FP_k}$, where: $K$ indicates the total number of classes, $TP_k$ denotes the number of samples that are actually class $k$ and are correctly predicted as class $k$.

The calculation of $Recall$ is analogous to that of $Precision$, involving computing it individually for each class and subsequently averaging the results.
$F1_k = 2 \times \frac{Precision_k\times Recall_k}{Precision_k+Recall_k}$, Subsequently, macro - averaging is performed on the $F1_k$, $F1 = \frac{1}{K}\sum_{k=1}^KF1_k$.

The results are presented in Table \ref{table1}.

\begin{table}[htp]
\centering
\caption{First Round of Detection Results.}
\begin{tabularx}{\columnwidth}{lXXXX}
    \toprule
    Model & Accuracy & Recall & F1 score & Precision \\
    \midrule
    GPT-4o & 55\% & 53\% & 59\% & 66\% \\
    \hline
    GPT-4Turbo & 41\% & 23\% & 32\% & 54\% \\
    \hline
    DeepSeek V3 & 58\% & 47\% & 52\% & 67\% \\
    \hline
    DeepSeek R1 & 56\% & 55\% & 55\% & 62\%\\
    \bottomrule 
\end{tabularx}
\label{table1}
\end{table}
   


\begin{figure}
\centering
\includegraphics[width=0.60\columnwidth]{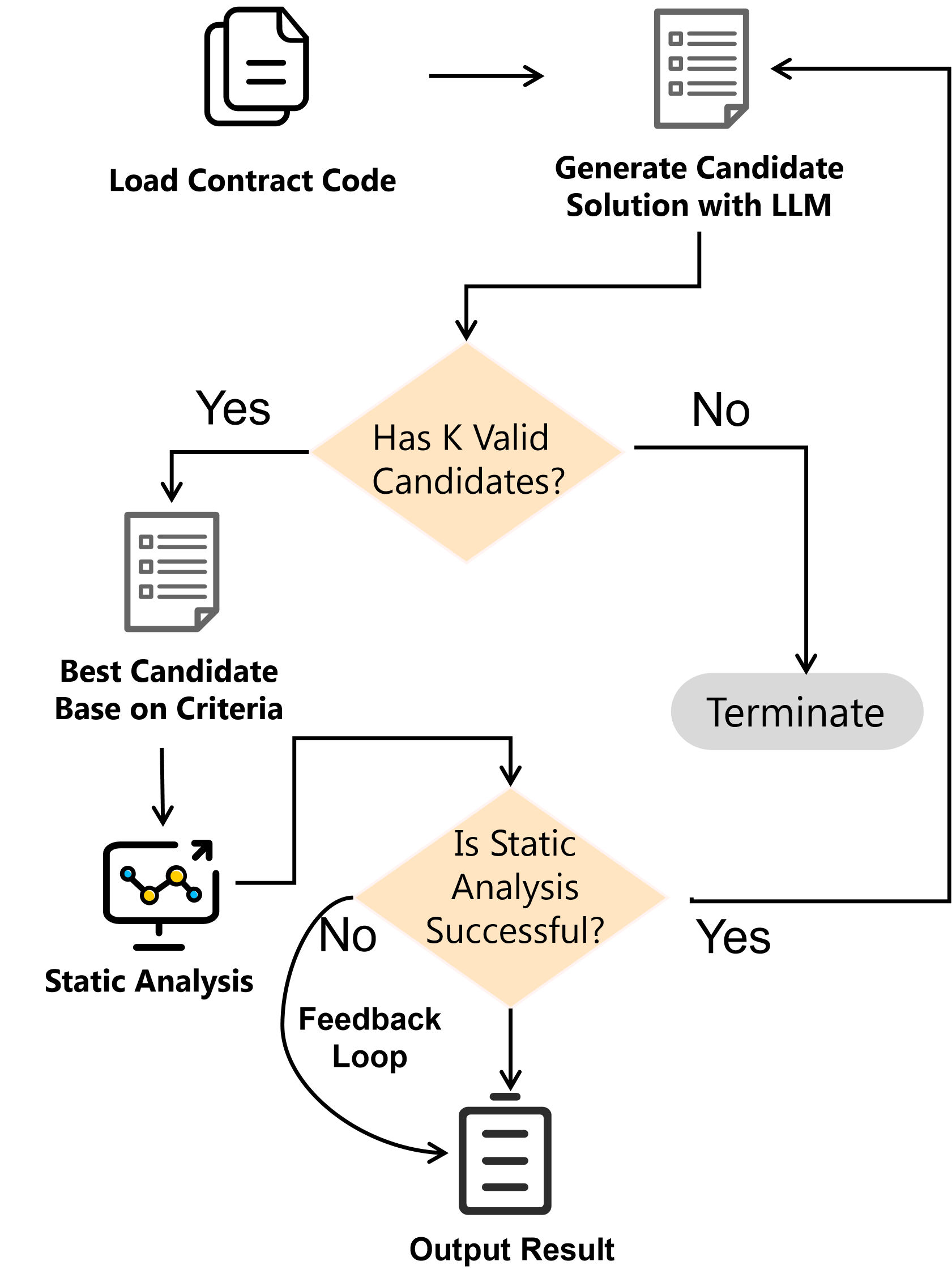} 
\caption{Workflow of the Loop Feedback Iterative Optimization Algorithm.}
\label{fig_loop}
\end{figure}

It can be inferred that, despite all models' superior logical reasoning and code comprehension capabilities, the evaluation metrics fell short of the acceptable criteria. This partly stems from the fact that certain library misuse patterns share similar code characteristics, potentially misleading GPT. Furthermore, the complexity of the contracts in the test data, likely attributed to their extensive codebase, compromised GPT's detection accuracy \citep{maity2024effective}. We implemented an iterative optimization method utilizing feedback loops to augment the GPT module's contract detection capabilities and diminish the false positive rate. As shown in Algorithm \ref{alg:algorithm1}. And Figure \ref{fig_loop} illustrates the workflow of the Loop Feedback Iterative Optimization Algorithm described in Algorithm \ref{alg:algorithm1}.

\begin{algorithm}[htb]
\caption{Loop Feedback Iterative Optimization Algorithm}
\label{alg:algorithm1}
\textbf{Input}: Smart contract code, Prompt($P_n$)\\
\textbf{Parameter}: Desired accuracy threshold, GPT model\\
\textbf{Output}: Final detected misuse pattern \\
\begin{algorithmic}[1]
\STATE Let $t=0$ denote the iteration count.
\WHILE{not Achieve Desired Accuracy}
\STATE Generate the prompt for the current iteration using $P_{t+1} = \text{GeneratePrompt}(t, P_t)$.
\STATE Provide the prompt to the GPT model and obtain the result: $R_{t+1} = \text{GPT.Analyze}(P_{t+1})$.
\IF {$R_{t+1}$ meets Accuracy Criteria}
\STATE The algorithm has converged; exit the loop.
\ELSE
\STATE Update the iteration count: $t \gets t + 1$.
\STATE Generate feedback based on the result: $F_{t+1} = \text{GenerateFeedback}(R_{t+1})$.
\STATE Update the prompt words with feedback for the next iteration: $P_{t+1} = \text{UpdatePromptWords}(P_t, F_{t+1})$.
\ENDIF
\ENDWHILE
\STATE \textbf{return} the final detected misuse pattern $R_{t+1}$.
\end{algorithmic}
\end{algorithm}

Specifically, the current GPT detection's output is used as an integral component of the subsequent detection round's prompt, aiming to induce a corrective feedback mechanism. The comprehensive procedure is outlined as follows:

\textbf{Definitions and Notations. }Let $C$ be the set of smart contract codes provided as input and let $P$ be the set of library misuse patterns, where $P=\{P1, P2,..., P8, NONE\}$ representing a total of 9 distinct patterns including a '$NONE$' category for no detected misuse. 

Let $G$ be the set of prompt words given to GPT, which includes the smart contract code, the library misuse patterns and their attributes, and the code snippets associated with each pattern.

Let $R$ represent the feedback from GPT's analysis, indicating which library misuse pattern the input contract corresponds to.

\textbf{Prompt Generation Function. }The function to generate the initial prompt, $GenerateInitialPrompt(C,P,G)$, can be expressed as Eq. \ref{equa:initialprompt}.
\begin{equation}
\label{equa:initialprompt}
    GenerateInitialPrompt(C,P,G) = C+P+G
\end{equation}
Let $P_n$ represent the prompt generated on the $n_{th}$ iteration

\textbf{Iterative Analysis Process.} First, we provide the initial prompt to GPT to obtain a preliminary result \(R_1\) indicating the detected library misuse pattern for the input $contract$ $C$. Then, we construct the feedback prompt shown in Eq. \ref{equa:feedbackpromt} that includes the initial analysis result and any discrepancies or negative feedback from the preliminary analysis. Finally, we repeat the analysis using the new prompt to refine the result and continue this iterative process until a satisfactory level of accuracy is achieved.
\begin{equation}
\label{equa:feedbackpromt}
    GenerateFeedbackPrompt(R_1) = R_1+NegativeFeedback(R_1)
\end{equation}


\subsection{Static Analysis Module of LibScan}
GPT employs logical reasoning to parse contract code by misuse pattern scenarios, yet it does not consistently concentrate on syntax or achieve precise matching of contract statements. In contrast to static analysis tools, which focus on particular code statements, no comparable tools currently exist to evaluate the LLM module's effectiveness in identifying library misuse within smart contracts. Nonetheless, leveraging a database from empirical research, replete with code snippets associated with diverse misuse patterns, has informed the development of our static analysis tool.

Specifically, we developed two distinct types of static analysis tools and selected the superior one based on a thorough evaluation of their performance. Subsequently, the selected tool's output was amalgamated with the findings from the LLM module to derive conclusive results:


\textbf{Based on Code Similarity Static Matching(BCSSM).} Utilize the TfidfVectorizer to transform code snippets and input code into a TF-IDF matrix \citep{jain2024effective}. TF-IDF, a text representation technique, accounts for term frequency (TF) and inverse document frequency (IDF), mitigating the influence of common words to enhance discernment of similarity. Compute the cosine similarity between the input code and individual code snippets. Cosine similarity quantifies the angular separation between two vectors, here represented by the TF-IDF-represented code snippets, with values approaching 1 signifying higher similarity.

\textbf{Hybrid Context-Sensitive Static Analysis(HCSA).} Based on the empirical research recommendations, the detection of patterns P1-P3 necessitates not only identifying potential vulnerabilities but also scrutinizing the coverage scope of library functions, while pattern P5 requires differential analysis between wrapped and primitive library functions. To address these four library misuse patterns, we eschew simplistic code-level matching at the static analysis stage. Instead, we leverage Slither \citep{feist2019slither}, a robust smart contract static analysis tool that converts Solidity contracts into Abstract Syntax Trees (ASTs) and constructs Control Flow Graphs (CFGs) \citep{du2025capturing} to facilitate contextual semantic analysis. This capability enables precise tracking of inheritance hierarchies, data dependencies, and function call semantics across contracts. For $P1$-$P3$ and $P5$, we implement customized detection rules through Slither's programmable interface, specifically targeting library override completeness ($P1$), arithmetic operation context validation ($P2$), modifier conflict resolution ($P3$), and wrapper-primitive function equivalence verification ($P5$). 

In contrast, patterns $P4$, $P6$-$P8$ remain detectable through TF-IDF matrix-based code-level matching inherited from BCSSM, as their identification primarily relies on structural similarity rather than contextual semantics.

\section{Experiment}
\subsection{Dataset.} We developed LibScan utilizing empirical study data obtained from \cite{huang2024revealing}. The dataset encompasses 1,018 smart contracts, which are from real-world projects that are currently operational and have been pre-labeled with library misuse patterns. We selected 662 samples as our test dataset, following the principles previously outlined. The study also furnished pertinent attribute descriptions and code snippets for the eight library misuse pattern types. Utilizing these code snippets, we identified code features and amalgamated them with attribute descriptions to formulate prompts for LLM.

\subsection{LLM Module.} 
\begin{figure}
\centering
\includegraphics[width=0.70\columnwidth]{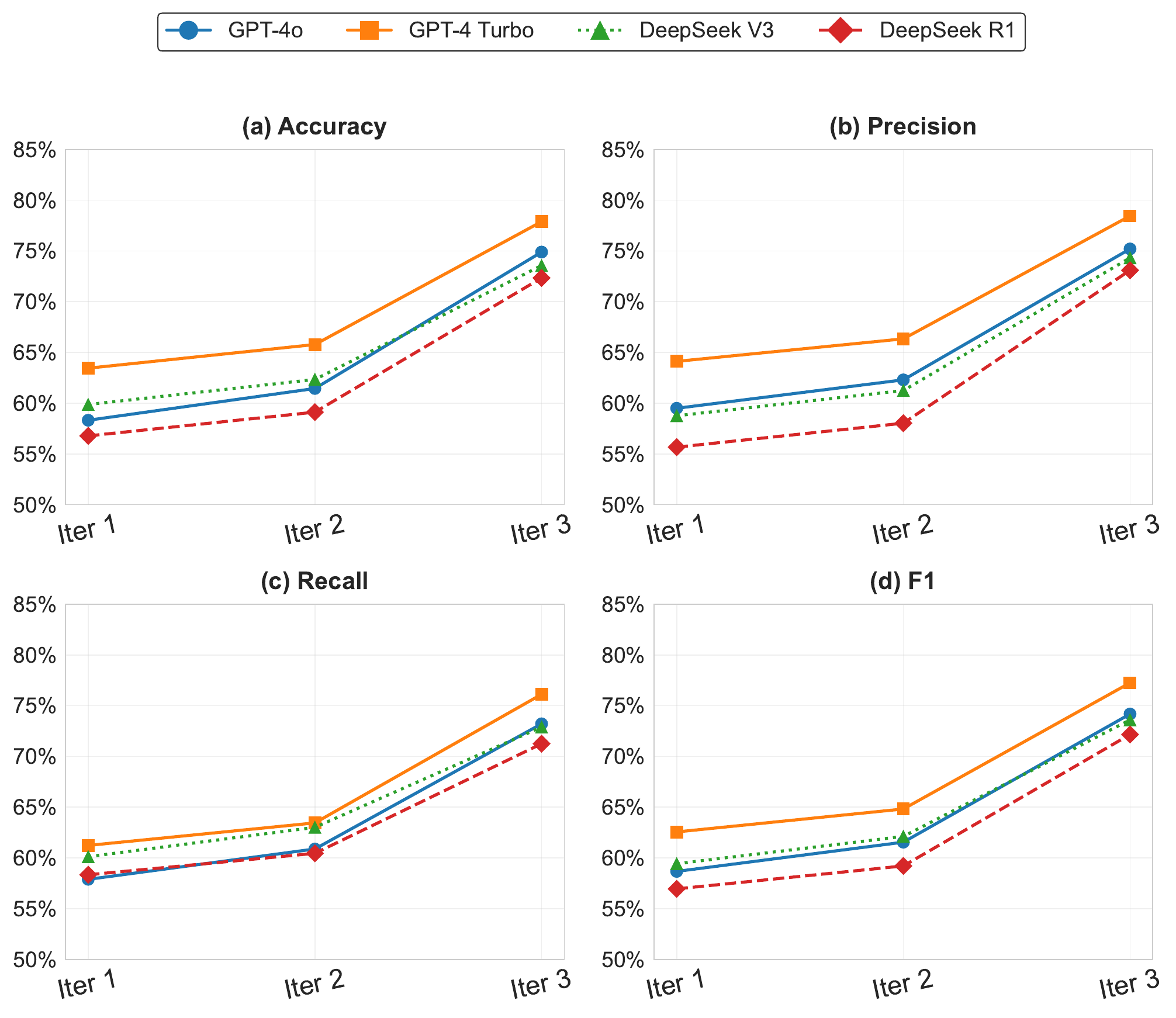} 
\caption{Four Models' Evaluation Metrics after Iteration}
\label{fig3}
\end{figure}

We customized a PTS prompt to enhance GPT's detection capabilities for misuse patterns in smart contract libraries and implemented an iterative optimization method incorporating feedback to diminish GPT's false positive rate in analyzing complex code. In our experimental trials, we evaluated GPT's output incorporating feedback, observing notable enhancements with each iteration, thereby validating the efficacy of our approach. Nonetheless, for DeepSeek V3 and GPT-4 Turbo, improvements plateaued after the third iteration, with an increase in erroneous predictions. Consequently, we selected the outcomes from the third iteration as definitive for all models. Figures \ref{fig3} illustrate the variation in the four evaluation metrics for all models across the three iterations.

\subsection{Static Analysis Module.} We have developed two specific static analysis tools aimed at detecting misuse patterns within smart contract libraries, identified as BCSSM and HCSA. We conducted tests on BCSSM and HCSA utilizing an identical dataset and subsequently computed the corresponding evaluation metrics for the outcomes, as illustrated in Table \ref{table2}.

\begin{table}[htp]
\centering
\caption{Evaluation Metrics for BCSSM and HCSA}
\begin{tabularx}{\columnwidth}{lXXXX}
    \toprule
    Method & Accuracy & Recall & F1 score & Precision \\
    \midrule
    BCSSM & 72.20\% & 72.91\% & 75.66\% & 79.38\% \\
    \hline
    HCSA & 70.66\% & 76.00\% & 76.29\% & 81.98\% \\
    \bottomrule
\end{tabularx}

\label{table2}
\end{table}

\subsection{Integration of Results Method.} Given the complex interdependencies between the erroneous outputs of the LLM and static analysis modules, and considering Random Forest (RF) \citep{eshghie2021dynamic} is well-suited to capturing such non-linear relationships, we adopted the RF model. The RF model integrates the output results of the LLM and static analysis modules, leveraging their complementary strengths to improve detection accuracy. 

RF is an ensemble learning technique that trains multiple decision trees and aggregates their outputs to produce class labels for classification or average prediction values for regression. It aggregates the prediction results of the LLM and static analysis modules using the voting mechanism of multiple decision trees. For an input sample, each decision tree independently predicts a class label, and the final result is determined by the class label with the highest number of votes, as shown below:
\begin{eqnarray}\label{eq:3}
\hat{y} = \underset{c \in \{p1,\ldots,p8,NONE\}}{\operatorname{argmax}} \sum_{i=1}^{N} \mathbb{I}(T_i(LLM, Static) = c)
\end{eqnarray}

where $T_i$ represents the prediction result of the i-th decision tree, and $\mathbb{I}$ is the indicator function.

The specific implementation steps are as follows:

\begin{itemize}
    \item \textbf{Feature Construction:}
        \begin{itemize}
            \item Two categorical features: LLM\_test (prediction results of the LLM module) and Sta\_test (prediction results of the static analysis module). Each feature is integer-encoded ($p1$→0, $p2$→1,..., $NONE$→8), with a total of 9 categories.
            \item Random Forest can directly handle categorical features, calculating the Gini impurity of category splits for splitting, eliminating the need for manual One-Hot encoding or other transformations.
        \end{itemize}
    \item \textbf{Training Phase:}
        \begin{itemize}
            \item Bootstrap sampling: Each decision tree is trained on a bootstrap sample of 1018 samples drawn with replacement to increase tree diversity and reduce model variance.
            \item At each node split, select the feature and threshold that minimize the Gini impurity, as shown in the formula:
            \begin{eqnarray}\label{eq:vcg}
            G(t) = 1-\sum_{c=1}^9 (p(c|t))^2
            \end{eqnarray}

            where $p(c|t)$ represents the proportion of samples in node t belonging to category c. $G(t)$ Measures category mixing in the node. If all samples are of the same category, then $G(t)=0$.
        \end{itemize}
    \item \textbf{Stopping Condition:} The pre-set max-depth (set to 10 in this module) or when the number of samples in a leaf node is less than min-samples-leaf (set to 5 in this module) is reached, to prevent overfitting and control model complexity.
    \item \textbf{Prediction Phase:} Each decision tree independently predicts a class label for new input samples. These predictions are then input into the function (\ref{eq:3}). The final result is the class label with the most votes from all trees.
    
\end{itemize}

\subsection{Evaluation}
In the experiments comparing four LLM module models, GPT-4 Turbo and DeepSeek V3 demonstrated superior performance in detecting misuse of 662 smart contract libraries (Figure \ref{fig3}). These two models had closely matched evaluation metrics.

We subsequently combined GPT-4 Turbo and DeepSeek V3 with the BCSSM and HCSA modules (from the static analysis experiment) using a random forest approach. Further analysis revealed that the GPT-4 Turbo-HCSA combination optimized contract assessment detection metrics. This combination served as the baseline for ablation experiments. As shown in Table \ref{table3}.



\begin{table}
\centering
\caption{Ablation Study Performance Comparison (Baseline: GPT-4 Turbo + HCSA), \textsuperscript{†} Baseline means the integration of LLM (GPT-4 Turbo) and static analysis (HCSA).}
\begin{tabularx}{\columnwidth}{lXXXX}  
\toprule
\textbf{Configuration} & \textbf{Accuracy (\%)} & \textbf{Recall (\%)} & \textbf{F1 Score (\%)} & \textbf{Precision (\%)} \\ 
\midrule
All\textsuperscript{†}  & 85.15 & 82.22 & 83.75 & 86.19 \\
HCSA only  & 70.66 & 76.00 & 76.29 & 81.98 \\
LLM only  & 73.56 & 72.89 & 74.32 & 73.60 \\
\bottomrule
\end{tabularx}
\label{table3}
\end{table}

Currently, aside from LibScan, no tools for detecting smart contract library misuse patterns have been identified. We selected Slither \citep{feist2019slither}, a widely-used static analysis tool, and developed customized detection rules for eight library misuse-pattern code snippets. Concurrently, we opted for GPTScan \citep{sun2024gptscan}, an LLM-based tool for detecting common smart contract vulnerabilities.

Adhering to its prompt framework design, we performed prompt reconstruction for the eight library misuse patterns to facilitate detection. We assessed these three tools using the same 662-sample smart contract test dataset, with the results presented in Table \ref{table4}.

\begin{table}
\centering
\caption{Performance Comparison of Tools.}
\begin{tabularx}{\columnwidth}{lXXXX}
\toprule
\textbf{Tool} & \textbf{Accuracy (\%)} & \textbf{Recall (\%)} & \textbf{F1 Score (\%)} & \textbf{Precision (\%)}\\ 
\midrule
LibScan & 85.15 & 82.22 & 83.75 & 86.19 \\
Slither & 62.39 & 67.85 & 64.17 & 60.24 \\
GPTScan & 68.91 & 63.52 & 71.68 & 66.73 \\
\bottomrule
\end{tabularx}
\label{table4}
\end{table}

As shown above, LibScan has demonstrated high accuracy in detecting smart contract library misuse patterns. This success can be attributed to its targeted parsing of these patterns, LLM prompt adjustments, and integration with a static analysis module to refine results.

\section{Related Work} 
Research related to this work spans three primary directions: (i) empirical studies of smart contract library misuse, (ii) smart contract vulnerability detection using static and dynamic analysis, and (iii) the application of large language models (LLMs) to code understanding and security analysis. This section situates LibScan within these research threads and highlights its distinctive contributions.

\textbf{Empirical Studies on Library Misuse in Smart Contracts. }Library misuse has long been recognized as a major source of software defects in traditional programming ecosystems. In the context of Ethereum smart contracts, this issue is exacerbated by immutability, economic incentives, and the difficulty of post-deployment patching. Prior empirical work has systematically investigated how developers misuse Solidity libraries and the resulting security implications.
\cite{huang2024revealing} conducted the first large-scale empirical study focusing specifically on library misuse in Ethereum smart contracts. Their work manually inspected real-world contracts and categorized misuse behaviors into eight recurring patterns, ranging from invalid wrapper checks to unnecessary library usage. This taxonomy provides a foundational understanding of how misuse manifests in practice and serves as a valuable dataset for subsequent research. However, their study primarily aimed at characterization rather than automation, and it did not propose a concrete detection tool capable of identifying these patterns in unseen contracts. Beyond Solidity-specific studies, prior research on API and library misuse in general-purpose languages has shown that misuse often arises from mismatches between developer assumptions and actual library semantics. These findings are consistent with the overestimated and underestimated library capability patterns observed in smart contracts, suggesting that library misuse is a cross-domain problem that requires semantic-aware analysis rather than purely syntactic checks.

\textbf{Static Analysis for Smart Contract Vulnerability Detection. }Static analysis remains one of the most widely adopted techniques for detecting vulnerabilities in smart contracts due to its scalability and determinism. Existing tools such as Slither \citep{feist2019slither}, SmartCheck \citep{tikhomirov2018smartcheck}, and SmartDagger \citep{liao2022smartdagger} analyze Solidity source code or EVM bytecode to identify known vulnerability patterns, including reentrancy, integer overflows, and access control flaws. These tools typically rely on predefined rules, abstract syntax trees, control-flow graphs, or data-flow analysis.
While effective for vulnerabilities with well-defined structural patterns, traditional static analyzers are less suitable for detecting library misuse. Many misuse patterns depend on nuanced semantic assumptions, partial function replacement, or incorrect expectations of library behavior, which are difficult to encode as rigid rules. For example, distinguishing between appropriate and inappropriate uses of the $using for$ directive often requires understanding the intended abstraction of the library rather than merely checking type compatibility.
Some recent work has explored hybrid static analysis approaches that incorporate contextual or inter-contract reasoning \citep{taseen2025hierarchical,pang2025graph} . These methods improve precision but still struggle with patterns that require implicit knowledge of library design intent or developer expectations. Consequently, static analysis alone is insufficient for comprehensive detection of library misuse patterns.

\textbf{Machine Learning Based Approaches. }To overcome the limitations of rule based analysis, several studies have proposed machine learning models for smart contract vulnerability detection \citep{huang2022smart,zhou2022vulnerability,crisostomo2025machine,hejazi2025comprehensive} . These approaches typically transform contract code into sequences, graphs, or embeddings and train classifiers to distinguish vulnerable from non-vulnerable contracts. Models based on recurrent neural networks, graph neural networks, and multi-task learning have demonstrated promising results for common vulnerability types.
However, most machine learning based approaches treat vulnerability detection as a coarse-grained classification problem and focus on a small set of well-known vulnerabilities. Library misuse, by contrast, often involves subtle deviations from best practices and does not always correspond to explicit security flaws. As a result, purely data-driven models may struggle to generalize without explicit pattern definitions and interpretability.

\textbf{Large Language Models for Code Analysis. } Recent advances in large language models have significantly improved the state of the art in code understanding, generation, and analysis. LLMs pretrained on large-scale code corpora have demonstrated strong capabilities in semantic parsing, logical reasoning, and cross-language transfer. These properties make LLMs attractive for tasks that require understanding developer intent and high-level semantics.
Several recent studies have explored the use of LLMs for smart contract security analysis. Tools such as GPTScan \citep{sun2024gptscan} employ LLMs to reason about contract logic and identify vulnerabilities, often in combination with lightweight program analysis. Empirical evaluations show that LLMs can capture complex vulnerability descriptions that are difficult to formalize as static rules. However, these approaches also report high false positive rates, particularly when analyzing large or complex contracts, due to hallucinations and overgeneralization. Other work has proposed prompt engineering strategies, chain-of-thought reasoning, and iterative feedback mechanisms to improve the reliability of LLM-based analyses. While these techniques reduce randomness and enhance reasoning transparency, they still require complementary mechanisms to validate LLM outputs against concrete code evidence.                             

\textbf{Hybrid Approaches Combining LLMs and Static Analysis. }Hybrid approaches that integrate LLM-based reasoning with static analysis have recently emerged as a promising direction. By delegating semantic understanding to LLMs and leveraging static analysis for precise code-level validation, such systems aim to balance expressiveness and soundness. Prior work in this area has primarily focused on detecting logic vulnerabilities or generating vulnerability explanations rather than identifying systematic misuse patterns.

LibScan differs from existing hybrid approaches in two key aspects. First, it targets library misuse patterns explicitly, grounding LLM reasoning in empirically derived pattern attributes and code-level scenarios. Second, it employs an iterative feedback mechanism and result integration strategy to mitigate false positives and reconcile discrepancies between LLM and static analysis outputs. To the best of our knowledge, LibScan is the first tool to operationalize an empirical library misuse taxonomy using an LLM–static analysis hybrid framework.

\section{Conclusion}
We introduce LibScan, a novel tool that integrates LLM with static analysis to detect library misuse patterns in smart contracts. By leveraging pattern attributes and iterative feedback optimization, LibScan effectively identifies eight misuse patterns while reducing false positives. Evaluations on 662 real-world contracts show its superior accuracy of 85.15\% and robustness, surpassing existing tools like Slither and GPTScan. Limitations include its ability to handle highly complex code semantics, which will be addressed in future work through enhanced model training and expanded pattern coverage. The open-source release of the tool aims to advance research and practice in smart contract security.

\section{Acknowledgments}
This work is sponsored by the National Natural Science Foundation of China (No.62362021 and No.62402146).


\printcredits

\bibliographystyle{cas-model2-names}

\bibliography{main}

\end{document}